\documentclass[12pt,preprint]{aastex}

\usepackage{amsfonts}
\usepackage{amssymb}
\usepackage{amsthm}
\usepackage{amsmath}
\usepackage{hyperref}

\shorttitle{Reflection of Radiation of a Rotating Source}
\shortauthors{Dementyev\,A.\,V.}

\begin{document}

\title{Specific Features of Reflection of Radiation of a Compact Source Rotating about its Axis}

\author{A. V. Dementyev\altaffilmark{1}}
\altaffiltext{1}{Department of Astrophysics, Faculty of Mathematics and Mechanics, Saint Petersburg State University (SPbSU), Universitetsky pr., 28, Stary Peterhof, Russia, 198504}
\email{a.dementiev@spbu.ru}

\begin{abstract}
Reprocessing of X-ray pulsar radiation by the atmosphere of a companion in a binary system may result in reflected pulse radiation with a period of the pulses equal to the period of the pulsar under suitable conditions.
In this paper the influence of the rotation of the source --- the pulsar --- about its axis on the parameters of thus reflected pulses is investigated. The binary system is modeled by the spherical reflective screen and the compact source uniformly rotating about its axis; the beam pattern (BP) of the source periodically runs over the surface of the screen. Irradiation of the screen by the pulses infinitely narrow in time and by the rectangular pulses is considered. In this model parameters of the pulses reflected in some directions are calculated. The main conclusion provided by the consideration of this model is that the properties of reflected pulses --- their profile, and the moments of reaching the observer --- substantially depend on the correlation between the light speed and the speed of the BP passing over the companion surface. The possibility of applying of the obtained results to the known X-ray accretion-powered pulsars and rotation-powered pulsars in binary systems is examined.

\end{abstract}


%%%%%%%%%%%%%%%%%%%%%%%%%%%%%%%%%%%%%%%%%%%%%%%%%%%%

\section{Introduction}

An important way of obtaining information about stars in binary systems is observation of the reflection effect which involves reprocessing and reemission of the radiant energy coming from a star by the atmosphere of another star. If a binary system includes a neutron star acting as a pulsar, with a companion occasionally getting into its beam pattern (BP), in this case some part of the incident flux of the radiant energy can be reemitted in the form of pulses with a period equal to the period of rotation of the neutron star --- the pulsar --- about its axis. The properties of the pulses thus reflected have a number of specific features to be considered in the present paper. 

It is known that reprocessing of of the X-ray pulsar radiation by the atmosphere of a companion may result in emergence of reflected pulses in the X-ray as well as ultraviolet and optical ranges of the spectrum \citep{BST1974}. Therefore, the objects for the study of which can be useful the results given below are, firstly,  X-ray pulsars emitting due to the accretion of the companion substance on them (accretion-powered pulsars), and, secondly, pulsars in binary systems emitting in the X-ray range due to their kinetic rotational energy (rotation-powered pulsars). Although these objects are obviously different in their physical characteristics and observational manifestations, it is their common property that is essential for us in this case: the presence of a compact X-ray source emitting in a limited solid angle $\Omega<4\pi$ and rotating like a lighthouse.

In these binary systems the reflection effect looks as follows. With appropriate orientation of the pulsar BP the companion side facing the pulsar is occasionally exposed to X-ray radiation pulses --- the BP runs over the surface of the companion. The interaction of this emission with the companion atmosphere results in reprocessing and reemission of X-ray photons getting into the atmosphere. Some part of this energy is spent on heating the atmosphere and increasing the stationary radiation flux from the companion. Another part of this energy can be reemitted in the form of pulses with the period equal to the period of pulsar rotation  both in the X-ray and longer-wavelength bands of the spectrum \citep{BST1974}. Besides, penetration of high-energy photons (as well as pulsar wind particles) into the companion atmosphere can result in evaporation of the companion in close binary systems \citep{FST1988, PEBK1988}. In this paper the formation of reflected pulse radiation is considered separately from other possible kinds of interaction between  the pulsar radiation and the companion.

The observation and interpretation of the reemitted pulses expand our possibilities of studying the properties of the source, companion and binary system in general \citep{AB1974}. It would seem, besides, that the study of a quick-changing radiation flux can provide more information than the study of a stationary flux: in the first case we have a time-dependent function while in the second case we have one constant value only (we abstract from the orbital movement and the companion rotation about its axis --- the periods of pulsars are usually considerably less than the typical time of these motions). An example of using observations of reemitted pulses is determination of the mass of the neutron star in Hercules~X-1 system \citep{MN1976}.

Unambiguous interpretation of the results of observations of reemitted pulses in the binary systems under consideration is likely to be a challenging task as it is necessary to consider a whole number of factors influencing the properties of pulses (we do not discuss the technical possibility of observation of such pulses here). Firstly, this is the characteristics of incident radiation: the kind of the spectrum and the beam pattern. Secondly, the structure of the companion atmosphere determining the way the reprocessing and reemission of the photons getting into the atmosphere proceed. The entering of a considerable quantity of high-energy photons and pulsar wind particles into the atmosphere results in its heating and, consequently, in its restructuring, which should be taken into account. Thirdly, radiation propagation can be influenced by the gas-dynamic processes existing in a binary system; it especially concerns X-ray pulsars. Finally, the properties of reemitted pulses depend on geometric parameters of the binary system.

The present paper considers the last, geometric, factor. It is considered on the basis of a simple model of the binary system that does not concern the details of the reprocessing and reemission of the photons. This model consists of a spherical reflective screen that does not emit anything as it is. This screen is exposed to the radiation of the source rotating about its axis. The size of the source allows considering it a point source. Due to the location of the reflective sphere all the points of its surface in the line-of-sight of the source occasionally get in the beam pattern of the source. Two kinds of the source BP are under consideration: in the first case of the BP every elementary area of the screen surface getting within its limits is exposed by a pulse having infinitely narrow time spread while in the second case it is irradiated by a rectangular pulse. It is supposed that every elementary area of the surface of the sphere instantaneously reemits all the radiation reaching it so that the brightness of the area appears to be equal in all directions. The observer recording the reflected pulses is at such a great distance from the screen-source system that the screen has point dimensions for him. The model is described in detail in the next section. 

The main conclusion provided by the consideration of this model is that the properties of reflected pulses --- their profile, and the moments of reaching the observer --- substantially depend on the correlation between the light speed and the speed of the BP passing over the companion surface. In particular, at a certain correlation between these speeds the maximum amplitude of the reflected pulses is achieved \citep{D2014}. Thus, it may be concluded that the rotation of the source irradiating the companion atmosphere can have a certain influence on the reflected pulse radiation. This influence, however, is determined by the geometric parameters of the binary system only and does not depend on the way the reprocessing and reemission of the photons getting into the atmosphere proceed.  

%%%%%%%%%%%%%%%%%%%%%%%%%%%%%%%%%%%%%%%%%%%%%%%%%%%%

\section{The model}

Let us specify the parameters of the screen -- source -- observer system described in the Introduction.

\begin{figure}
\centerline{
\includegraphics[height=6.0cm,width=15cm,angle=0]{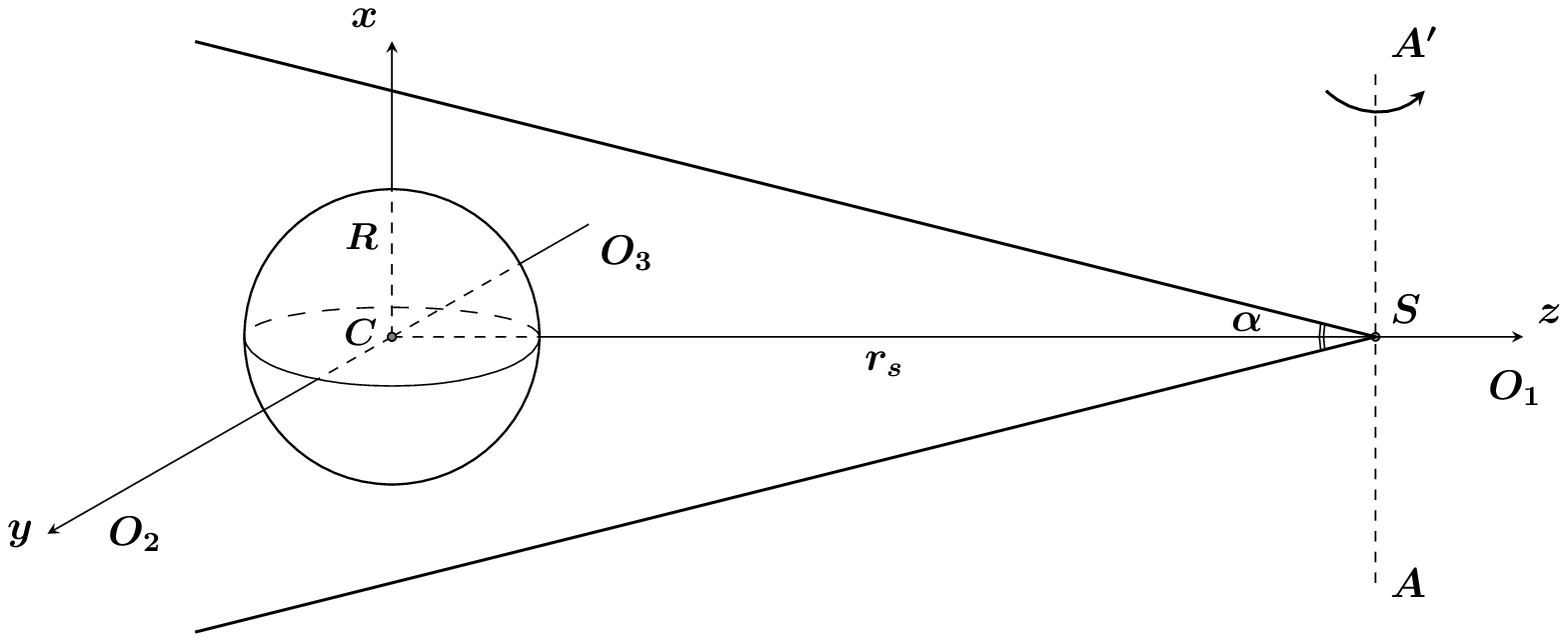}}
\caption{Scheme of arrangement of the spherical screen with the center in point $C$ and radius $R$, the point source $S$ and beam pattern of the source --- plane angle $\alpha$. The distance between the source and the center of the sphere is designated with $r_s$. Line $AA'$ is the axis of the rotation of the source; the arc with an arrow shows the direction of its rotation.
}\label{EkranIstochnik}
\end{figure}

{\it{Screen.}} The reflective screen is spherical with radius $R$. The sphere has no radiation of its own. We will assume the following reflecting properties of the surface of the sphere: every elementary area of the surface of the sphere instantaneously reemits all the radiation falling on it so that its brightness appears to be the same in all directions (lambertian source). We will relate the Cartesian coordinate system $(x, y, z)$ with the sphere placing the beginning of this system in its center $C$ (Fig. \ref{EkranIstochnik}). We will also introduce the spherical coordinate system $(r, \theta, \varphi)$ with the center in point $C$:
\begin{equation}\label{spheric_coordinate}
x=r\sin{\theta}\cos{\varphi}\,,\quad y=r\sin{\theta}\sin{\varphi}\,,\quad z=r\cos{\theta}\,.
\end{equation}

{\it{Source.}} Let the linear size of the radiation source be such as compared to other typical dimensions of the system that it may be considered a point source. The source is in $\theta=0$ direction (i.e. on $Cz$ axis) at the distance of $r_s$ from the center of the sphere so that
\begin{equation}\label{r_s_gg_R}
r_{s}=kR\,,\quad\mbox{где} \quad k\gg1\,.
\end{equation}
 We will designate the location of the source with $S$. The source rotates about its axis $AA'$ parallel with $Cx$ with the constant angular velocity of $\omega=2\pi/P$ where $P$ is the period of rotation. The source rotation is counterclockwise if we look at the plane $(yz)$ from the side of the positive direction of $Cx$ axis (Fig. \ref{EkranIstochnik}).

Regarding the beam pattern of the source we will first assume the following. We will consider that the source radiation is concentrated within the limits of the plane angle $\alpha$ with the vertex in $S$ point that remains perpendicular to the plane $(yz)$. In addition, if we place the angle $\alpha$ in the plane $(xz)$, the cross-section of the sphere by this plane will be completely within $\alpha$ (Fig. \ref{EkranIstochnik}). We will also consider that within the BP limits the source emits uniformly. We neglect the dilution of radiation during its propagation from the source at the distance of $r_s-R$ to the point away from the source at the distance of $r_s$.

{\it{Observer.}} The radiation reflected by the sphere is recorded at the distance $r_o$ from the center of the sphere, where
\begin{equation}\label{r_o_gg_r_s}
r_{o}\gg r_{s}\gg R\,,
\end{equation}
so the sphere is a point source for the observer. We will be interested in the points of observation lying in three possible directions: $CO_1\,(\theta=0)$, $CO_2\,(\theta=\pi/2;\, \varphi=\pi/2)$, $CO_3\,(\theta=\pi/2;\, \varphi=3\pi/2)$ --- all of them located in the plane $(yz)$. The transition from one point to another may be interpreted as observation of the binary system in different phases of its orbital motion from one point. For the purposes of brevity these directions will be further designated as $O_1$, $O_2$ and $O_3$, respectively. 

We will be interested only in the relative changes of the radiation flux in the point of observation during the period abstracting from the properties of the radiation itself (spectral composition, absolute flux value, etc.) as well as from the details of the source emission mechanism, radiation reprocessing by the reflective surface and radiation propagation in the space. Thus, the parameters of the problem are $r_s$, $R$ and $P$, their correlation influencing the properties of the reflected pulses recorded by the observer. 

As the BP of the source is a plane angle oriented as stated above, the radiation reaches every elementary area of the surface of the sphere getting into the BP in the form of pulses having infinitely narrow time spread, their period being equal to $P$ due to the source rotation. As the radiation does not reach different areas simultaneously and the distances from the areas to the observation points are not equal, the observer records the radiation reflected from the sphere already in the form of pulses of finite duration as well as recurring with the period $P$ \citep{AB1974}.

{\it{Note 1.}} It is usually conceived that BP of a pulsar consists of two identical beams symmetrically located relative to the pulsar rotation axis. In this case we could also take a point source with the BP consisting of two identical plane angles symmetrical relative to the source rotation axis. Within the frames of our model this would just mean that the screen is exposed to radiation with the period of $P/2$. Therefore, it is sufficient to confine ourselves to consideration of the source with the BP consisting of one beam.    

{\it{Note 2.}} Consideration of the BP in the form of a plane angle means that the sphere is irradiated by pulses, their profile in time being described by the Dirac $\delta$-function. Let the profile of the reflected pulses be defined by the pulse function $h(\tau)$ here. If the source BP is such that a pulse with a certain $s(\tau)$ profile rather than a $\delta$-like pulse falls on every area of the sphere surface, the profile of the reflected pulses $\phi(\tau)$ is found by convolution of $h(\tau)$ and $s(\tau)$ \citep{AB1974}
\begin{equation}\label{svertka}
\phi(\tau)=\int_{-\infty}^{+\infty} h(\tau')\,s(\tau-\tau')\,d\tau'\,.
\end{equation}
 Use of Equation (\ref{svertka}) supposes that separate $\delta$-like pulses are reflected from the screen surface independently from one another, i.e. the reflection process is linear. In fact, the purpose of the present paper is to obtain the pulse function $h(\tau)$ for some cases. After calculation of $h(\tau)$ using Equation (\ref{svertka}) we will find $\phi(\tau)$ for the case of rectangular incident pulses. 

%%%%%%%%%%%%%%%%%%%%%%%%%%%%%%%%%%%%%%%%%%%%%%%%%%%%

\section{Specific features of formation of the reflected pulses}

Let us consider the screen--source system section by the plane $(yz)$. We will introduce the Cartesian $(\tilde x, \tilde y)$ and polar $(\tilde r, \tilde{\varphi})$
\begin{equation}\label{polar_coordinate}
\tilde x=\tilde r\cos{\tilde{\varphi}}\,,\quad \tilde y=\tilde r\sin{\tilde{\varphi}}
\end{equation}
coordinate system in this plane with the center in point $S$. We will direct the $S\tilde{x}$ axis contrarily to the $Cy$ axis and the $S\tilde y$ axis contrarily to the $Cz$ axis. The locus of $(\tilde x \tilde y)$ plane  or, what is the same, $(yz)$ plane which were reached by the radiation at some moment of time $t$ is the Archimedean spiral determined by equation
\begin{equation}\label{spiral_equation}
\tilde r=c(t-\tilde{\varphi}/\omega)\,,
\end{equation}
where $c$ is the speed of light \citep{BG1972}. Spiral (\ref{spiral_equation}) is actually the section of the wave front by $(\tilde x \tilde y)$ plane.
\begin{figure}
\centerline{
\includegraphics[height=5.7cm,width=15cm,angle=0]{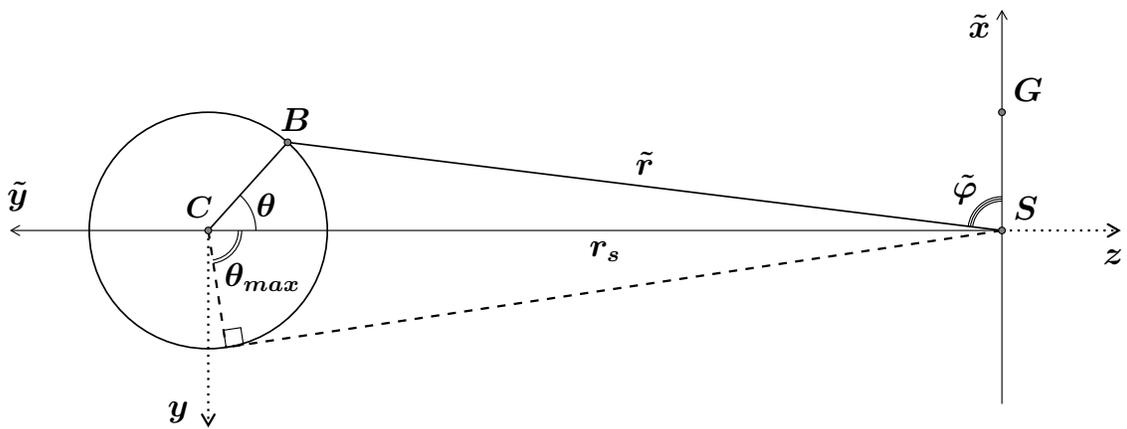}}
\caption{Section of the screen--source system by $(yz)$ plane. Here $CB=SG=R$.
}\label{EkranIstochnik_yz}
\end{figure}

Let us consider two characteristic time intervals. The first of them is the time of passage of the distance equal to the sphere diameter by radiation:
\begin{equation}\label{time_of_raduis}
t_1=2R/c\,.
\end{equation}
The second of them is the full time of BP passage over the surface of the sphere, i.e. the time of the source turn by angle
\begin{equation}\label{def_beta}
2\arcsin{R/r_s}=2\arcsin{1/k}\simeq2/k
\end{equation}
(see Fig. \ref{EkranIstochnik_yz}) where condition (\ref{r_s_gg_R}) is taken into account. This time is equal to
\begin{equation}\label{time_of_turn}
t_2=\frac{2/k}{2\pi}\,P=\frac{P}{\pi k}\,.
\end{equation}
 Below we will measure the time in the units of the period $P$ and the distance in the units of the sphere radius $R$  and will designate the dimensionless time through $\tau$. We will also introduce the $W=cP/R$ parameter which represents the dimensionless speed of radiation propagation.

\begin{figure}
\centerline{
\includegraphics[height=6.0cm,width=9.5cm,angle=0]{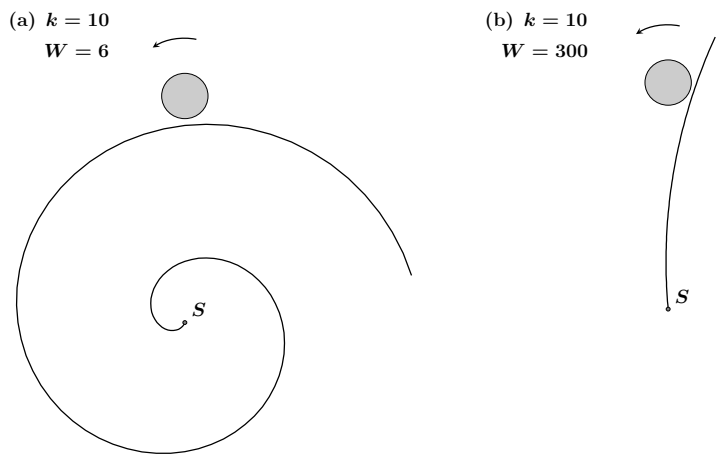}}
\caption{Wave front section by $(yz)$ plane with two values of $W$ parameter. The grey circle at the top of every figure is the screen section by the same plane. The arcs with arrow show the direction of rotation of the source located in point $S$.}\label{spiral}
\end{figure}

Let $t_2\ll t_1$, which means $W\ll2\pi k$. Then the arm of the spiral (\ref{spiral_equation}) is twisted  at the distance $r_s$ from the source (Fig. \ref{spiral}a). In this case it may be considered that a plane wave front falls on the screen and passes over the screen with the speed of light in the direction opposite to the direction of $Cz$ axis (Fig.~\ref{EkranIstochnik}).

In the contrary case when $t_2\gg t_1$, we have $W\gg2\pi k$. In this case the spiral arm just starts twisting at the distance $r_s$ from the source (Fig. \ref{spiral}b). The picture of the screen irradiation can be represented as follows. The plane wave front passes over the screen in the direction of $Cy$ axis (Fig. \ref{EkranIstochnik}), i.e. from right to left in accordance to Fig. \ref{spiral}b. The speed of motion of this front is $v=\omega r_s$; in dimensionless values $V=2\pi k$. The direction of propagation of the radiation itself is perpendicular to the direction of the front motion.

In case of intermediate values $t_2\sim t_1$ and, accordingly, $W\sim2\pi k$, the wave front passing over the screen can be probably no longer considered a plane one and its motion is no longer rectilinear.

Thus, two extreme modes of radiation of the sphere can be identified depending on the correlation between two characteristic time intervals $t_2$ and $t_1$ (we can also speak about a correlation between two speeds: $W$ --- the speed of light and $V$ --- the speed of passage of the source BP over the surface of the screen). It should be expected that the reflected pulses reaching the observer will have different parameters depending on the particular screen irradiation mode.  

\begin{figure}
\centerline{
\includegraphics[height=8.0cm,width=9cm,angle=0]{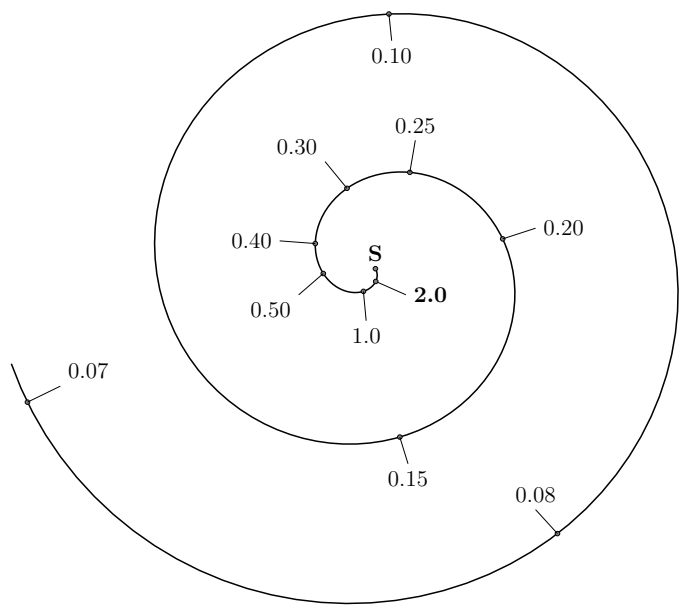}}
\caption{Spiral (\ref{spiral_equation}) on which the possible points of its crossing with the center of the reflective sphere for different values of $Q$ parameter are plotted. The word ''possible'' means the assumption that the wave front passes through the center of the sphere (through point $C$) ''without noticing'' its surface; actually the radiation cannot get inside the sphere, of course. The numbers near the respective points show the $Q$ values for which the center of the sphere would be located in the point marked. In particular, for Figure \ref{spiral}a the parameter $Q=0.10$ and for Figure \ref{spiral}b $Q=4.77$.}\label{spiral_Q}
\end{figure}

{\it{Note.}} The particular mode of irradiation of the spherical screen does not depend on its radius $R$. It is actually determined by the way the dimensionless $Q$ value
\begin{equation}\label{Q_def}
 Q\equiv\frac{W}{V}=\frac{cP}{2\pi r_s}
\end{equation}
 correlates with the unit (Fig. \ref{spiral_Q}). However, the parameters of the reflected pulses with the same $Q$ are different depending on the $R$ value. Firstly,  the bigger the radius of the sphere, the larger the reflective surface; therefore, the pulse amplitude may also appear larger. Secondly, the pulse duration equal by order of magnitude $t_1+t_2$ also depends on $R$ according to Equations (\ref{time_of_raduis}) and (\ref{time_of_turn}).

%%%%%%%%%%%%%%%%%%%%%%%%%%%%%%%%%%%%%%%%%%%%%%%%%%%%

\section{Method of calculation of the parameters of the reflected pulses}

We will consider that individual reflected pulses do not overlap. In this case it is sufficient to calculate the change of the flux of the reflected radiation in the time interval equal to one period. Let us consider the small area $d\sigma$ located on the reflective sphere so that the polar angle of the spherical coordinate system of the place of its location is equal to $\theta$ and the azimuthal angle is equal to $\varphi$. For the areas that can be reached by the radiation of the source, the $\theta$ angle changes from $0$ to
\begin{equation}\label{theta_max}
\theta_{max}=\arccos{(R/r_s)}=\arccos{(1/k)}
\end{equation}
(Fig. \ref{EkranIstochnik_yz}) and the $\varphi$ angle --- from $0$ to $2\pi$. We can represent the contribution of the radiation reflected by the area $d\sigma$ over the period to the total flux recorded by the observer using the $\delta$-function as follows:
\begin{equation}\label{potok_d_sigma}
 h_o\,\delta(\tau-\tau_{so})\cos{\psi_s}\cos{\psi_o}\,d\sigma\,. 
\end{equation}
Here $h_o$ is the radiation flux over the period that would be created by the area in the point of observation if it were oriented perpendicular both to the direction towards the source and to the line of sight; $\tau_{so}=\tau_{so}(\theta,\,\varphi)$ is the moment of time when the radiation from this area is recorded by the observer. The $\psi_s$ angle is the angle of the radiation incidence on the area under consideration and the observer sees this area at $\psi_o$ angle.

Considering that $d\sigma=R^2\sin{\theta}\,d\theta\,d\varphi$, the full $h(\tau)$ flux reflected from the semisphere facing the observer will be equal to
\begin{equation}\label{F_total}
h(\tau)=h_o\,R^2\int\nolimits^{\varphi_{max}}_{\varphi_{min}}d\varphi\int\nolimits^{\theta_{max}}_{0}\delta(\tau-\tau_{so})\cos{\psi_s}\cos{\psi_o}\sin{\theta}\,d\theta\,,
\end{equation}
where $\varphi_{min}$ and $\varphi_{max}$ are the limits of change of the azimuthal angle on this semisphere. This formula means the following. To find the value of $h$ flux at some moment of time $\tau$ it is necessary to integrate the integrand taken without the $\delta$-function only for the values of  $\theta$ and $\varphi$ with which we have $\tau_{so}(\theta,\,\varphi)=\tau$ (see in detail below).

\begin{figure}
\centerline{
\includegraphics[height=4.8cm,width=16cm,angle=0]{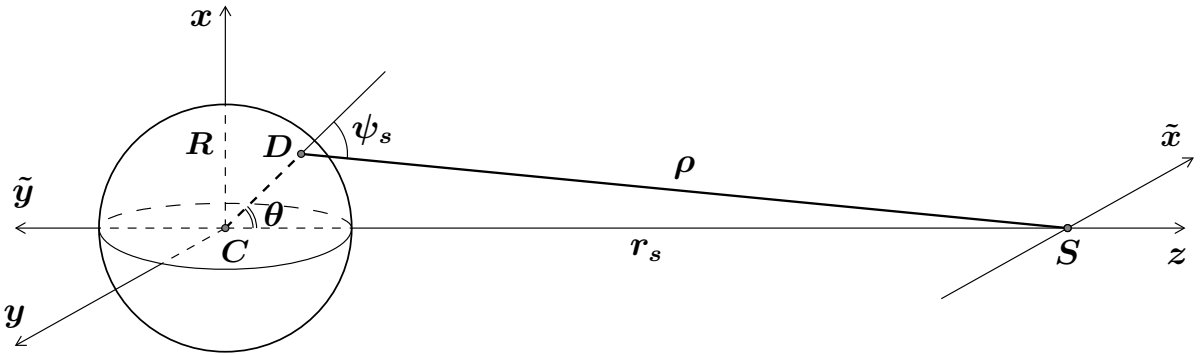}}
\caption{Identification of $\psi_s$ --- angle of radiation incidence on the elementary area of the screen surface.}\label{Ugol_Psi}
\end{figure}

{\it{Identification of $\psi_s$ angle.}} The angle of radiation incidence on the area is the angle between the normal to the area and the direction towards the source (Fig. \ref{Ugol_Psi}). Using the law of cosines we get
\begin{equation}\label{cos_psi_s}
\cos{\psi_{s}}=\frac{r_{s}\cos{\theta}-R}{\rho}=\frac{r_{s}\cos{\theta}-R}{\sqrt{r_{s}^2-2r_{s}R\cos{\theta}+R^2}}=\frac{k\cos{\theta}-1}{\sqrt{k^2-2k\cos{\theta}+1}}\,,
\end{equation}
 where the third equation follows from Equation (\ref{r_s_gg_R}).

{\it{Identification of $\psi_o$ angle.}} Let the observer be located in $O_1$ direction. This observer sees the area under consideration at $\psi_{o1}$ angle, its cosine being identified similarly to $\cos{\psi_s}$:
\begin{equation}\label{cos_psi_o1}
\cos{\psi_{o1}}=\frac{r_{o}\cos{\theta}-R}{\sqrt{r_{o}^2-2r_{o}R\cos{\theta}+R^2}}\approx\cos{\theta}\,,\quad (\theta\in[0; \, \theta_{max}],\enskip  \varphi\in[0; \, 2\pi])\,.
\end{equation}
The approximate equation here is obtained with account of condition (\ref{r_o_gg_r_s}). Given in parentheses are limits of change of $\theta$ and $\varphi$ of the areas reflecting the radiation that are in the direct line of sight for observer $O_1$.

For observers in $O_2$ and $O_3$ directions we similarly have
\begin{equation}\label{cos_psi_o2}
\cos{\psi_{o2}}\approx\sin{\theta}\sin{\varphi}\,,\quad (\theta\in[0; \, \theta_{max}], \, \varphi\in[0; \pi])
\end{equation}
and
\begin{equation}\label{cos_psi_o3}
\cos{\psi_{o3}}\approx-\sin{\theta}\sin{\varphi}\,,\quad (\theta\in[0; \, \theta_{max}], \enskip \varphi\in[\pi; 2\pi])\,.
\end{equation}

{\it{Identification of the moment $\tau_{so}$.}}  Let $t_s$ be the delay of the moment of radiation incidence on some area $d\sigma$ relative to the moment of incidence on the area for which $\theta=0$ (this area is the nearest to the source). The radiation reflected by the area $d\sigma$ will also reach the observer with a certain delay $t_o$ relative to the radiation reflected by the area nearest to the source. Both $t_s$ and $t_o$ may appear either larger or less than $0$. Thus, the radiation passes the distance
\begin{equation}\label{path_O1}
(r_s-R)+c(t_s+t_o)+(r_o-R)\,,
\end{equation}
from the source to the observer in the direction $O_1$ and
\begin{equation}\label{path_O2_3}
(r_s-R)+c(t_s+t_o)+r_o
\end{equation}
from the source to the observer in the direction $O_2$ or to the observer in the direction $O_3$. There is a specific middle term in both cases for each area. At the same time the sought-for moment of time expressed in the units of the period interesting for us is
\begin{equation}\label{tau_so}
\tau_{so}=\tau_s+\tau_o\,.
\end{equation}

First let us get the expression for $\tau_o$. As we suppose that the observer is at infinitely long distance from the screen, the direction towards the observer from all $d\sigma$ areas are parallel to one another. For the observer in the direction of $O_1$
\begin{equation}\label{tau_o_O1}
\tau_o\equiv\tau_{o_1}=\frac{d}{cP}=\frac{R}{cP}\,(1-\cos{\theta})=\frac{1}{W}\,(1-\cos{\theta})\,,
\end{equation}
where $d$ is the length of the perpendicular dropped from the point with $(R,\theta,\varphi)$ coordinates on the plane perpendicular to $Cz$ axis and passing through the point on this axis for which $z=R$. Similarly, for observers in the directions of $O_2$ and $O_3$ we have
\begin{equation}\label{tau_o_O2_O3}
\tau_o\equiv\tau_{o_2}=\tau_{o_3}=-\,\frac{1}{W}\,|\sin{\theta}\sin{\varphi}|\,.
\end{equation}

To determine the $\tau_s$ delay we must compare the moments of radiation incidence on the relevant areas of the sphere surface. We will measure off the time from the moment when the source BP projection on the plane $(\tilde x \tilde y)$ is the semiaxis $S\tilde x$. In this case the moment of radiation incidence on the area located in the neighbourhood of some point $B$ of the plane (see Fig. \ref{EkranIstochnik_yz}) is equal
\begin{equation}\label{moment_B}
t_B=\tilde r/c+\tilde{\varphi}/\omega\,.
\end{equation}
Note that the reference time may be any other moment: then in the right part of Equation (\ref{moment_B}) there will appear some constant as another summand. But as we just need the difference between the moment of radiation incidence on the areas, this constant will reduce anyway in making up this difference.

If the elementary area $d\sigma$ is located in the neighbourhood of some point $D$ not lying in the $(\tilde x \tilde y)$ plane (see Fig. \ref{Ugol_Psi}), the moment of radiation incidence on this area is
\begin{equation}\label{moment_D}
t_D=\rho/c+\tilde{\varphi}_D/\omega\,,
\end{equation}
where $\tilde{\varphi}_D$ angle is measured off from the $S\tilde x$ axis to the projection of $\rho$ segment on the $(\tilde x \tilde y)$ plane. Let point $D$ have the Cartesian coordinates $(R\sin{\theta}\cos{\varphi},R\sin{\theta}\sin{\varphi},R\cos{\theta})$. In accordance with Equation (\ref{moment_D}) the time of radiation propagation from the source --- point $S$ to the $d\sigma$ area --- point $D$ is
\begin{equation}\label{moment_D_tau}
\tau_D=\frac{\rho}{R\,W}+\frac{\tilde{\varphi}_D}{2\pi}\,,
\end{equation}
where
\begin{equation}\label{rho}
\rho=\sqrt{r_s^2-2r_sR\cos{\theta}+R^2}=R\,\sqrt{k^2-2k\cos{\theta}+1}\,.
\end{equation}
We will determine the $\tilde{\varphi}_D$ angle from the triangle lying in $(yz)$ plane, its vertexes being: 1) the point of location of the source $S$; 2) the projection of point $D$ on the $(yz)$ plane; its coordinates are $(0,R\sin{\theta}\sin{\varphi},R\cos{\theta})$; 3) some point on the positive $S\tilde x$ semiaxis ; we will take point $G$ with the Cartesian coordinates $(x,y,z)$ equal to $(0,-R,r_s)$ as this point (Fig. \ref{EkranIstochnik_yz}). The sought-for angle is the angle at $S$ vertex. According to the law of cosines we have
\begin{equation}\label{cos_tilde_varphi}
\cos{\tilde{\varphi}_D}=\frac{-\sin{\theta}\sin{\varphi}}{\sqrt{k^2-2k\cos{\theta}+\cos^2{\theta}+\sin^2{\theta}\sin^2{\varphi}}}\,.
\end{equation}
We will designate the moment of radiation incidence on the elementary area of the sphere surface nearest to the source through $\tau_p$ (for it $\theta=0$). With account for Equations (\ref{moment_D_tau}--\ref{cos_tilde_varphi}) this moment is equal to
\begin{equation}\label{tau_p}
\tau_p=\frac{k-1}{W}+\frac{1}{4}\,.
\end{equation}
Thus,
\begin{equation}\label{tau_s}
\tau_s=\tau_D-\tau_p=\frac{\sqrt{k^2-2k\cos{\theta}+1}-k+1}{W}+\frac{\tilde{\varphi}_D}{2\pi}-\frac{1}{4}\,,
\end{equation}
where $\tilde{\varphi}_D$ is found as an arccosine of the expression (\ref{cos_tilde_varphi}).

{\it{Calculation of the values of $h(\tau)$.}} The values of the $h(\tau)$ function determined by the Equation (\ref{F_total}) can be obtained as follows. On the intervals of $\theta$ and $\varphi$ integration as well as on the interval of $\tau_{so}$ change we will select some sets of discrete points
\begin{equation}\label{uzly_theta}
0=\theta_1<\theta_2<\dots<\theta_i<\dots<\theta_{q_1-1}<\theta_{q_1}=\theta_{max}\,,
\end{equation}
\begin{equation}\label{uzly_phi}
\varphi_{min}=\varphi_1<\varphi_2<\dots<\varphi_j<\dots<\varphi_{q_2-1}<\varphi_{q_2}=\varphi_{max}\,,
\end{equation}
\begin{equation}\label{uzly_tau}
\tau_{min}=\tau_1<\tau_2<\dots<\tau_n<\dots<\tau_{q_3-1}<\tau_{q_3}=\tau_{max}\,,
\end{equation}
where $\tau_{min}$ and $\tau_{max}$ are the moments of arrival of the rising and falling edges of the pulse, respectively, and $q_1$, $q_2$, and $q_3$ are some integer numbers that may also be identical. The $\tau_{min}$ and $\tau_{max}$ are initially unknown to us; therefore, we will select an interval of change $\tau_{so}$ so that $\tau_{min}$ and $\tau_{max}$ should get to the interval {\it{a priori}}. After the calculation procedure of $h(\tau_n)\equiv h_n$ values the $\tau_{min}$ and $\tau_{max}$ moments are defined as the first and last moment of time with which $h_n\ne0$. It is convenient to take the points (\ref{uzly_theta}--\ref{uzly_tau}) uniformly with steps $\Delta\theta$, $\Delta\varphi$, and $\Delta\tau$ respectively.

The procedure of calculating the $h(\tau)$ flux values starts from assigning the value equal to $0$ to $h_n$ values for all $n$. Then $\tau_{so}$ is calculated for every pair of points $\theta_i$ and $\varphi_j$ by the Equations (\ref{tau_so}--\ref{tau_o_O2_O3}, \ref{tau_s}) and the interval $(\tau_{n-1};\,\tau_n]$ is found so that $\tau_{n-1}<\tau_{so}\leqslant\tau_n\,$. After that the
\begin{equation}\label{F_tau_k_value}
h_o\,R^2\,\frac{\cos{\psi_s(\theta_i)}\,\cos{\psi_o(\theta_i\,,\varphi_j)}\,\sin{\theta_i}\,\Delta\varphi\,\Delta\theta}{\Delta\tau}
\end{equation}
value is added to the current $h_n$ value.  Division by $\Delta\tau$ was needed here because the described procedure is actually numerical integration for interval $(\tau_{n-1};\,\tau_n]$ and we refer the value thus received to one point, namely to $\tau_n$.

{\it{Calculation of the values of the convolution.}} After finding the pulse function values $h_n$  we can find the $\phi(\tau)$ profile of the reflected pulses resulting from irradiation of the sphere with pulses with arbitrary $s(\tau)$ profile using Equation (\ref{svertka}). The $\phi(\tau_k)$ convolution values are calculated as usual \citep{B2000}:
\begin{equation}\label{svertka_sum}
\phi(\tau_k)=\sum_{n=1}^{q_3}h_ns_{k-n}\,.
\end{equation}
As an example we will consider the rectangular profile of incident pulses
\begin{equation}\label{s_rect}
s(\tau)=\begin{cases}
1/a\,,\quad0\leqslant\tau\leqslant a<1 \\
0\,,\quad\quad\tau<0\,,\tau>a\,.
\end{cases}
\end{equation}

%%%%%%%%%%%%%%%%%%%%%%%%%%%%%%%%%%%%%%%%%%%%%%%%%%%%

\section{Irradiation of the sphere by parallel rays}

Let us assume that the sphere is at an infinitely large distance from the source. Then all the elementary areas on the sphere are exposed to radiation by parallel rays and for all of them $\tilde{\varphi}_D=\pi/2$. With such a mode of irradiation of the sphere the spiral arm (\ref{spiral_equation}) is twisted at the distance $r_s$ from the source ($Q\ll1$). In this case instead of Equations (\ref{theta_max}) and (\ref{cos_psi_s}) we have
\begin{equation}\label{theta_max_approx}
\theta_{max}=\pi/2
\end{equation}
and
\begin{equation}\label{cos_psi_s_approx}
\cos{\psi_{s}}=\cos{\theta}
\end{equation}
respectively. Then obviously
\begin{equation}\label{taus=tauo} 
\tau_s=\tau_{o_1}
\end{equation}
for the observer in the $O_1$ direction. With account for Equation (\ref{tau_o_O1}) we have
\begin{equation}\label{tau_so_O1_approx}
\tau_{so}=\frac{2}{W}\,(1-\cos{\theta})\,.
\end{equation}
In this case
\begin{equation}\label{tau_so_O1_approx_granicy}
0\leqslant\tau_{so}\leqslant\frac{2}{W}\,.
\end{equation}
Thus, the observer will record the radiation that the sphere reflects over the period during the time $2/W$. In other words, the $2/W$ value is the duration of the reflected pulse, with the pulse incident on the sphere having infinitely narrow time spread. 

For the observer to see individual not overlapping reflected pulses their duration must be less than the period, i.e.
\begin{equation}\label{period}
\frac{2}{W}\leqslant1\quad\mbox{or}\quad\frac{2R}{c}\leqslant P\,.
\end{equation}
The $P/(2R/c)$ value is the duty cycle of the reflected pulses. 

The expression (\ref{tau_so_O1_approx}) for $\tau_{so}$ allows obtaining $h(\tau)$ in an explicit form. Namely, in the integral (\ref{F_total}) we will consider that $\varphi_{min}=0$, $\varphi_{max}=2\pi$, will substitute $\cos{\psi_o}$, $\theta_{max}$ and $\cos{\psi_s}$ from the Equations (\ref{cos_psi_o1}), (\ref{theta_max_approx}) and (\ref{cos_psi_s_approx}), respectively, and make the following substitution of the variable: $u\equiv2(1-\cos{\theta})/W$. Then
\begin{align}\label{F_O1_approx}
h(\tau)&=2\pi h_oR^2\,\frac{W}{2}\,\int\nolimits^{2/W}_{0}\left(1-\frac{W}{2}\,u\right)^2\delta(\tau-u)du=\\
&=2\pi f_oR^2\cdot\left(\frac{W}{2}\right)^3\cdot(2/W-\tau)^2 \,.\nonumber
\end{align}
Here we used the well-known property of the $\delta$-function. With $2/W\leqslant\tau\leqslant1$ the flux $h(\tau)=0$. For observers $O_2$ and $O_3$ in this case of irradiation of all the areas of the sphere with parallel rays we fail to get an explicit expression for $h(\tau)$ in the same way as $\tau_{so}$ appears to depend both on $\theta$ and $\varphi$. 

In accordance with Equation (\ref{F_O1_approx}) the amplitude of the rising edge of the reflected pulse in the point of observation is
\begin{equation}\label{F_loc_1(0)}
h(0)=\pi h_oR^2\,W=2\pi h_oR^2\,\frac{P}{(2R/c)}
\end{equation}
--- the bigger the $R$ radius of the sphere, the larger the reflecting surface; thus, the larger the amplitude of the pulses. The $h_o$ value is the radiation flux over the period. The entire $h_o$ flux reaches the observer with $2R/c$ time that is less than or equal to the period in accordance with Equation (\ref{period}). The decreasing duty cycle of the pulses with unchanging $h_o$ flux and $P$ period results in the fact that all this flux will take less time and, hence, the amplitude of the pulse is supposed to grow, which is shown by the Equation (\ref{F_loc_1(0)}). According to Equation (\ref{period}) the pulse duration depends on $R$. This means that the moments of arrival of the rising and falling edges of the pulse to the observer also depend on $R$. 

The parameters of the reflected pulses obviously depend on the sphere radius $R$ both for the considered case of the irradiation of the sphere by parallel rays when $Q\ll1$ and for other possible values of $Q$. Besides, the influence of $R$ value on the parameters of the pulses takes place both for the observer in the $O_1$ direction and for any other directions in which the reflected pulses are recorded.  

%%%%%%%%%%%%%%%%%%%%%%%%%%%%%%%%%%%%%%%%%%%%%%%%%%%%

\section{Results}

\begin{figure}
\centerline{
\includegraphics[height=22.0cm,width=15cm,angle=0]{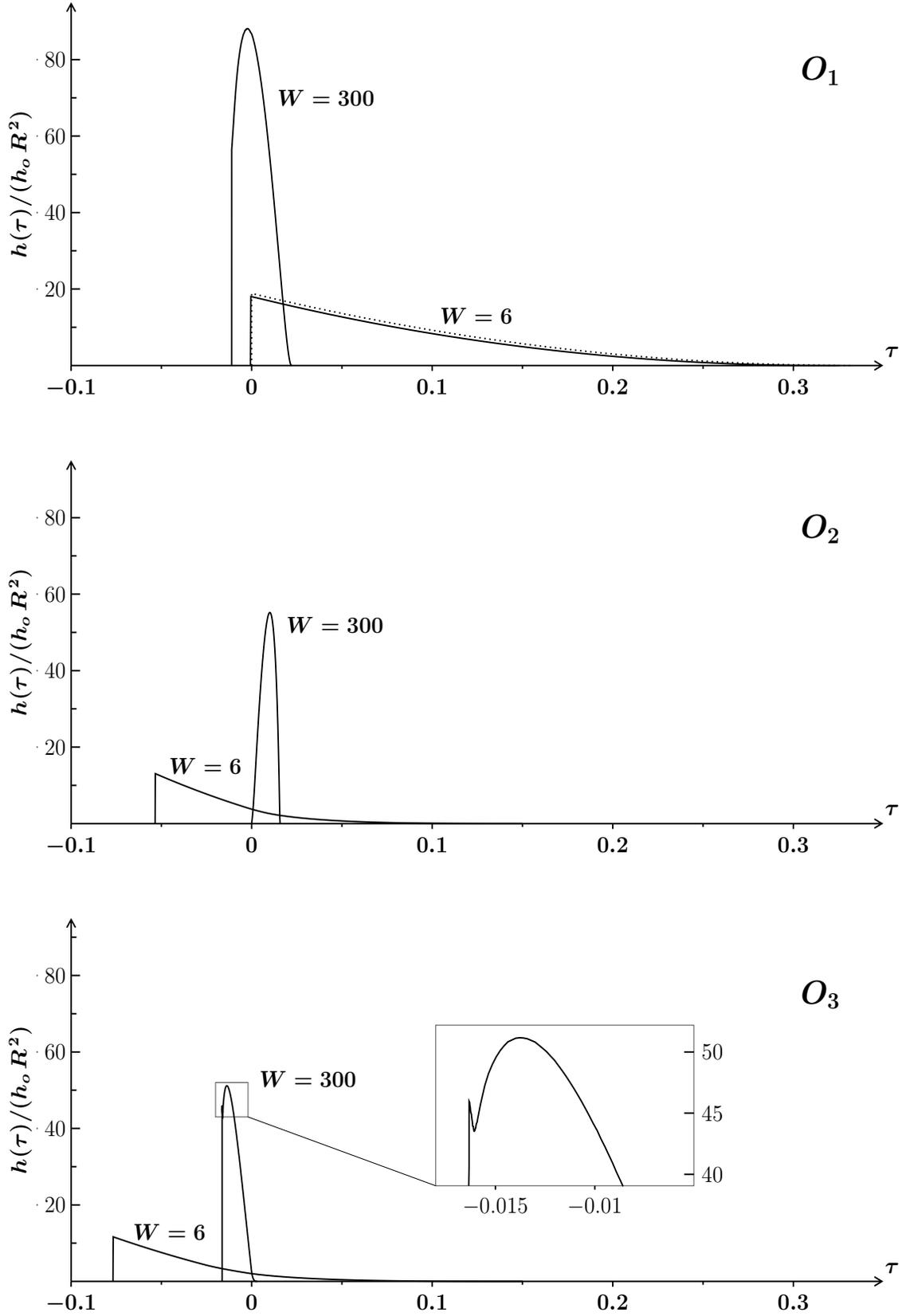}}
\caption{Profiles of the pulses reflected in $O_1$, $O_2$ and $O_3$  directions ($k=10$).}\label{impuls_Oall}
\end{figure}

Fig. \ref{impuls_Oall} presents profiles of the pulses reflected in $O_1$, $O_2$ and $O_3$ directions; the profiles have been plotted for two values of $W$ parameter that match the different modes of the irradiation of the sphere (see Fig. \ref{spiral}). Moment $\tau=0$ on the time scale corresponds to arrival of the radiation reflected by the sphere surface area that is the nearest to the source (for it $\theta=0$). In the upper figure showing the pulses recorded by the observer in $O_1$ direction the dotted line gives the pulse profile plotted according to Equation (\ref{F_O1_approx}) with $W=6$.  An unexpected specific feature of the profile of the pulse recorded in $O_3$ direction with $W=300$ is its nonmonotonic growth --- there is a small peak on the curve (see the lower figure). 

Note that for $O_2$ and $O_3$ directions the areas under the curves with any $W$ values are the same and approximately equal to $0.553$. For $O_1$ direction the areas under the curves are also identical with any $W$ and equal approximately $1.926$. The area under the dotted line in the upper figure is somewhat larger than this value as larger cosines of the angle of incidence are taken in the integrand in Equation (\ref{F_total}). 
Integration of the expression (\ref{F_O1_approx}) shows that this area is equal to $2\pi/3\approx2.094$ for any $W$.

The Table \ref{tbl-1} presents the following parameters of the reflected pulses: $[h/(h_oR^2)]$ --- the maximum value of the pulse profile (pulse amplitude); $\tau_b$ --- the moment of arrival of the pulse rising edge at the observer; $\tau_{max}$ --- the moment at which the pulse profile reaches the maximum value; $\Delta\tau_{1/2}$ --- the width of the pulse measured at the level of the half of its amplitude. The values of these parameters are given for the $O_1$, $O_2$ and $O_3$ directions being considered; for all cases $k=10$. The line at $W=6$ with values in bold face shows the corresponding values for the pulse profile plotted according to the expression (\ref{F_O1_approx}).

Let us discuss the obtained results. The profile of the reflected pulses is determined by elementary areas of the sphere surface which the observer sees at the moment. Figures \ref{impuls_Oall} and the data of the Table \ref{tbl-1} show that a change of the sphere irradiation mode (change of the correlation between $W$ and $V$) has a substantial impact on the profile of the reflected pulses as well as on the moments of their arrival at the observer. Let us imagine that $W$ has the value $\sim1$ and then $W$ starts growing due to the growing period of rotation of source $P$ with the unchanged radius of the sphere $R$, the distance between the source and the sphere $r_s$ and full source radiation flux. The amplitude of the reflected pulses will increase up to a certain maximum value exceeding the initial value by more than an order; in this case the duration of pulses decreases. This increase of the pulse amplitude occurs only through redistribution of the moments of arrival of radiation of different areas at the observer. Having reached the maximum the pulse amplitude slowly decreased for the same reason.

Another important conclusion from the received results is the following. Rotation of the source irradiating the sphere results in the situation that the picture of the reflected pulses looks different for $O_2$ and $O_3$ observers. In case of the untwisted spiral arm (Fig. \ref{spiral}b) the pulses arriving at these observers differ both by the profile shape and the arrival moments; in case of the twisted spiral arm (Fig. \ref{spiral}a) --- by the arrival moments only. In this case the difference between the moments of arrival of the pulses at $O_2$ and $O_3$ observers in accordance with Equations (\ref{cos_tilde_varphi}) and (\ref{tau_s}) is determined by $W$ and $k$ values (the delay $\tau_o$ is probably the same for both observers, Eq.(\ref{tau_o_O2_O3})). The data of the table show that at the values of $W$ corresponding to the mode of the twisted spiral arm the difference between the moment of pulse arrival at these observers does not depend on $W$ and is equal approximately to 0.023 (with $k=10$). Thus, knowing this difference between the pulse arrival moments we can find the companion radius $R$ using the known $P$ and $r_s$. In all probability using the pulse arrival moments for other directions (apart from $O_1$, $O_2$ and $O_3$) we can also obtain both $R$ and $r_s$ using the known period $P$.

{\it{Note 1.}} Note that if $k$ increases at constant $W$, the difference between the moments of pulse arrival at $O_2$ and $O_3$ observers disappears. This actually means that the source rotation stops influencing the properties of the reflected pulses. If $O_2$ and $O_3$ observers recorded the radiation of the pulsar in the Crab Nebula reflected by the Earth, they would obtain an identical picture of reflected pulses. 

\begin{figure}
\centerline{
\includegraphics[height=8.0cm,width=13cm,angle=0]{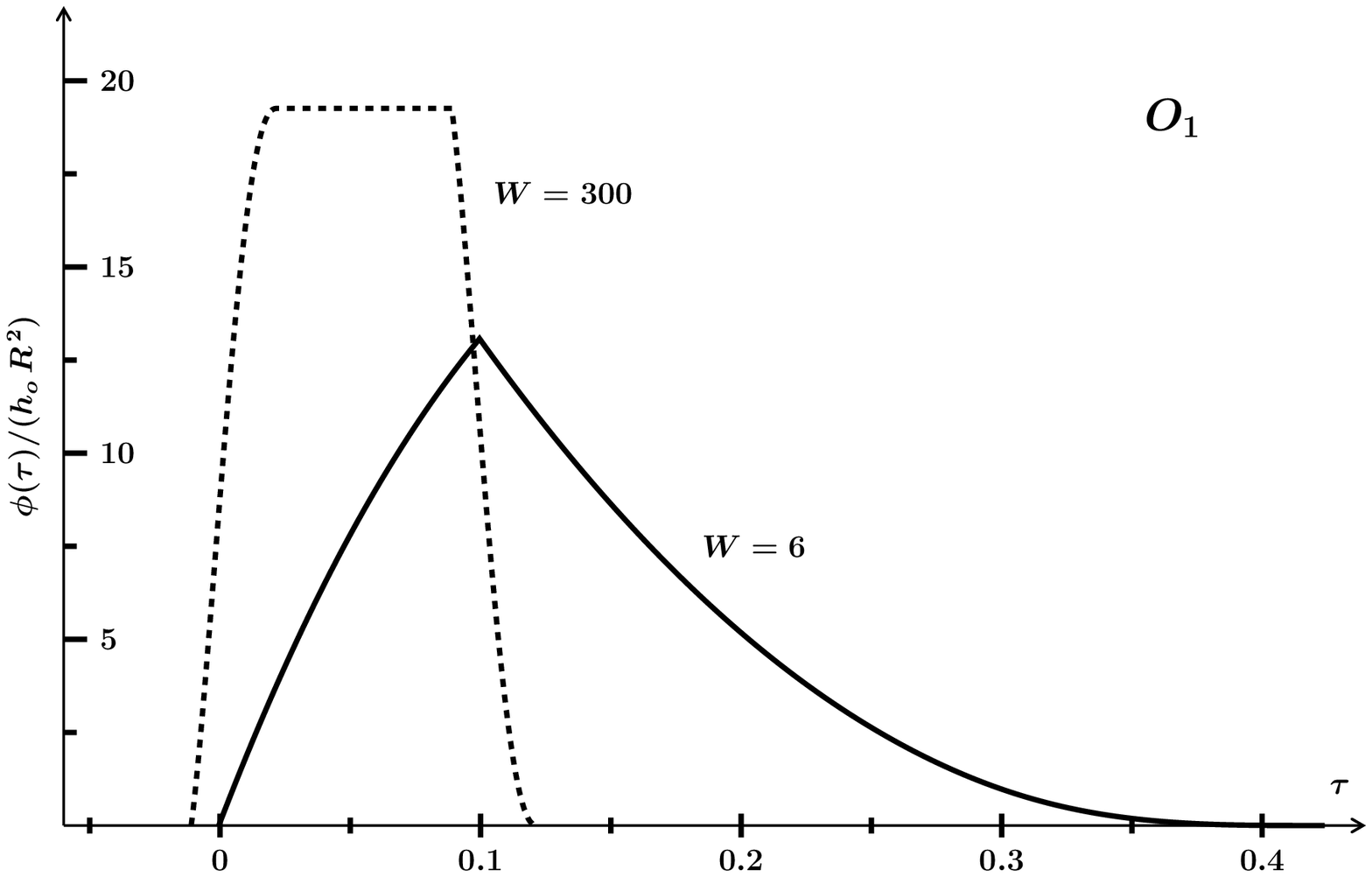}}
\caption{Profiles of the pulses reflected in $O_1$ direction with exposure of the sphere to radiation with rectangular pulses with the duration $a=0.1$ ($k=10$). }\label{impuls_O_rectangle}
\end{figure}

{\it{Note 2.}} The spreading of the pulses incident on the sphere surface is evidently to result in leveling of the differences between the properties of reflected pulses with different irradiation modes. For example, Fig. \ref{impuls_O_rectangle} presents profiles of the pulses reflected in $O_1$ direction with the rectangular profile of the incident pulses; the duration of the incident pulses was taken as follows: $a=0.1$.
\begin{deluxetable}{rrrrrrrrrrrrrrrrr}
%\begin{center}
\tabletypesize{\footnotesize}
\rotate
\tablecolumns{17} 
\tablewidth{0pc} 
\tablecaption{Parameters of the reflected pulses. \label{tbl-1}} 
\tablehead{ 
\colhead{}    & \colhead{}    & \multicolumn{3}{c}{$[h/(h_oR^2)]_{max}$} &  \colhead{}   &
\multicolumn{3}{c}{$\tau_b$} &  \colhead{}   & \multicolumn{3}{c}{$\tau_{max}$} &  \colhead{}   & \multicolumn{3}{c}{$\Delta\tau_{1/2}$}\\ 
\cline{3-5} \cline{7-9}\cline{11-13}\cline{15-17}\\ 
\colhead{$W$} & \colhead{$Q$} & \colhead{$O_1$}   & \colhead{$O_2$}    & \colhead{$O_3$} &\colhead{} & 
 \colhead{$O_1$}  & \colhead{$O_2$}   & \colhead{$O_3$}  & \colhead{} &  \colhead{$O_1$}  & \colhead{$O_2$}   & \colhead{$O_3$}    & \colhead{} & \colhead{$O_1$}  & \colhead{$O_2$}   & \colhead{$O_3$}}    
\startdata 
3&0.05 & 9.2 & 6.5  &6.1&&-0.000&-0.118&-0.141&&-0.000&-0.118&-0.141&&0.183&0.073&0.077  \\
6 &0.10& 18.2 & 13.3&11.8 && -0.000&-0.053&-0.077 &&-0.000&-0.053&-0.077&&0.093&0.036&0.040       \\
 \bf{6}& \bf{0.10} &  \bf{18.8} &\nodata &\nodata&&   \bf{0} &\nodata&\nodata&& \bf{0}&\nodata&\nodata&&   \bf{0.098}&\nodata& \nodata      \\
10 &0.16& 30.2 &22.7 &18.8 && -0.001&-0.028&-0.051&&-0.001&-0.028&-0.051&&0.062&0.021&0.025       \\
20 &0.32& 58.1 &46.1 &33.3 && -0.002&-0.009&-0.032&&-0.002&-0.009&-0.032&&0.029&0.010&0.013       \\
30 &0.48& 82.3 &62.6 &44.0&&-0.002&-0.003&-0.026&&-0.002&-0.003&-0.026&&0.020&0.008&0.010       \\
35 &0.56& 92.6 &64.6 & 48.1&&-0.003 &-0.002&-0.024&&-0.003&-0.002&-0.024&&0.018&0.008& 0.009      \\
40 &0.64& 101.5 &60.5 &51.6 &&-0.003 &-0.001&-0.023&&-0.003&-0.001&-0.023&&0.016&0.010&0.008       \\
63 &1.00&  127.3&56.0 & 61.2&& -0.004&-0.000&-0.020&&-0.004&0.006&-0.020&&0.013&0.011&0.008       \\
85 &1.35& 133.5 & 58.8&64.4 &&-0.006 &-0.000&-0.018&&-0.006&0.007&-0.018&&0.012&0.010&0.008       \\
90 &1.43& 133.4 & 59.4& 64.7&&-0.006 &-0.000&-0.018&&-0.006&0.008&-0.018&&0.013&0.010&0.008       \\
100 &1.59& 131.4 &60.0 &64.8&&-0.006&-0.000&-0.018&&-0.006&0.008&-0.018&&0.013&0.010&0.008       \\
150 &2.39&110.1  &58.6 &61.4 &&-0.008 &-0.000&-0.017&&-0.008&0.009&-0.017&&0.018&0.010&0.009       \\
300 &4.77& 90.0 &55.7 &51.5&&-0.011&-0.000&-0.016&&-0.002&0.010&-0.013&&0.023&0.011&0.011       \\
3000  &47.7& 89.1  &53.4 &53.0&&-0.015&-0.000&-0.016&&-0.000&0.012&-0.012&&0.024&0.011&0.011       \\
\enddata 
%\end{center}
\end{deluxetable} 

%%%%%%%%%%%%%%%%%%%%%%%%%%%%%%%%%%%%%%%%%%%%%%%%%%%%

\section{Application to astronomical objects}

As it was mentioned in the Introduction, the objects to which the results obtained in the present work can be applied are X-ray accretion-powered pulsars as well as rotation-powered pulsars in binary systems. The reference to the relevant catalogs \citep{LPH2006,LPH2007,ATNF,MHTH2005,CKKS1996,K2005} shows that the observed values $P$ and $r_s$ of the majority of these systems give $Q\ll1$, i.e. irradiation of the companion with pulsar radiation corresponds to a twisted spiral arm (Fig. \ref{spiral}a). But there are objects for which this is not so. Let us consider some corresponding examples formally calculating parameter $Q$ for them using the known values of $P$ and $r_s$. We are not discussing the real possibility of emergence of reflected pulse radiation in these systems.

The examples of interest for us among the high-mass X-ray binary systems \citep{LPH2006} are, in particular, objects 2S~0114+650, 4U~1538-52, and IGR~J16320-4751. 

{\it{2S~0114+650.}} The pulsar period in this system is $P=9605$~s \citep{BF2005}. Using the information given by \citet{RCC1996} ---  $P_{orb}\simeq11.6$~days, the mass of the neutron star $M_{NS}\simeq1.7\,M_{\Sun}$, the companion mass $M_c\simeq16\,M_{\Sun}$, the companion radius $R\simeq2.6\cdot10^{12}$~cm, and using the 3rd Kepler's Law
\begin{equation}\label{3_Kepler_law}
r_s=\left(\frac{GP_{orb}^2(M_{NS}+M_c)}{4\pi^2}\right)^{1/3}\,,
\end{equation}
we can find that for this object $r_s\simeq3.9\cdot10^{12}$~cm, $Q\simeq11.8$, and $k\simeq1.5$ ($G$ --- gravity constant).

{\it{4U~1538-52.}} According to \citet{MKHN1987} in this system $P\simeq530$~s, $P_{orb}\simeq3.7$~days, $M_{NS}\simeq1.8\,M_{\Sun}$, $M_c\simeq16\,M_{\Sun}$, $R\simeq9.3\cdot10^{11}$~cm. Therefore, for this object $r_s\simeq1.8\cdot10^{12}$~cm, $Q\simeq1.4$, and $k\simeq1.9$.

{\it{IGR~J16320-4751.}} In this   system the pulsar period is $P=1303$~s \citep{RBK2006} while the orbital period is $P_{orb}=8.96$~days \citep{CBB2005} . Considering the fact that the pulsar companion is a supergiant of the spectral type BN0.5Ia \citep{CCZ2013}, we assume $M_{NS}+M_c\simeq20\,M_{\Sun}$. Then we obtain $r_s\simeq3.4\cdot10^{12}$~cm and $Q\simeq1.8$.

Among the low-mass X-ray binary systems \citep{LPH2007} the highest $Q$ value is obtained for the object 4U~1626-67. According to \citet{MMNW1981}, $P\simeq7.7$~s, $P_{orb}\simeq2485$~s, $M_{NS}+M_c\simeq1.9\,M_{\Sun}$, $r_s\simeq3.4\cdot10^{10}$~cm. Therefore, $Q\simeq1.1$. \citet{LMM1988} have evaluated the radius of the companion as follows: $R\simeq5.6\cdot10^{9}$~cm; hence, $k\simeq6.1$.

For all the currently known rotation-powered pulsars in binary systems  \citep{ATNF,MHTH2005} we get $Q\ll1$. The highest $Q$ value is provided by the parameters of the binary pulsar J0737-3039B: $P\simeq2.8$~s, $P_{orb}\simeq2.4$~hours, $M_{NS}+M_c\simeq2.6\,M_{\Sun}$  \citep{LBK2004}; in this case $Q\simeq0.15$. The next biggest $Q$ value is obtained for the pulsar B1718-19: with $r_s\simeq1.4\cdot10^{11}$~cm \citep{LBHB1993} we have $Q\simeq0.034$.

Real binary systems with a pulsar are certainly far from the simple model considered by us. However, if the conditions in the binary system allow for emergence of reflected pulse radiation, it is necessary to take into account the mode of companion irradiation while interpretating the observations of such radiation (Fig.~\ref{spiral}).  Besides, the dependence on the irradiation mode will affect the parameters of the reflected pulses despite the finite time of processing of the incident radiation in the companion atmosphere, the ellipsoidal shape of the companion, the non-Euclidicity of the space in the neighbourhood of the neutron star and other possible differences of the binary system from our model. 

%%%%%%%%%%%%%%%%%%%%%%%%%%%%%%%%%%%%%%%%%%%%%%%%%%%%

\section{Conclusion}

The papers considering reprocessing and reemission/reflection of pulse radiation usually assume that due to the geometrical factor the reemitted pulses have greater duration compared to the incident ones by the order of $L/c$, where $L$ is the linear size of the reflective screen, see, for example, \citet{BST1974,AB1974}. For our model this time is $t_1$, see Equation (\ref{time_of_raduis}). The main conclusion of the present work is that if the irradiation of the screen with pulse radiation occurs due to the source rotation, it is also necessary to consider the increase of the duration of the reemitted pulses by a value equal to the time of passage of the source beam pattern over the surface of the screen --- time $t_2$ in our model, see Equation (\ref{time_of_turn}). If $Q\ll1$ (the mode of the twisted spiral arm Fig.~\ref{spiral}a), the reflected pulse spreading due to the geometrical factor is fully determined by time $t_1$ only. In the opposite case, when $Q\gg1$ (the mode of the untwisted spiral arm Fig. \ref{spiral}b), this spreading is determined by time $t_2$ only. A significant fact is that in intermediate cases the reflected pulse duration may appear substantially lower than $t_1+t_2$. Moreover, the minimal possible duration of the pulses appears to be less both than $t_1$ and $t_2$. The decrease of the duration of the pulses with other conditions being equal results in growth of its amplitude.

Another conclusion is that with any $Q$ the conditions of the pulsar irradiation of the right and left (from the point of view of $O_1$ observer) semispheres of the companion are different. This provides a principally new opportunity to determine the pulsar companion radius in the binary system and the distance between the pulsar and the companion by analyzing the parameters of the pulses reflected in different directions, which can be done by observing the companion in the course of its orbital motion.

\acknowledgments

The author is grateful to G.\,M.\,Beskin (SAO RAS) for drawing attention to the processes of interaction of radiation of the pulsars with their companions in binary systems. The author is also appreciative of V.\,V.\,Ivanov (SPbSU) for discussion of geometrical specific features of the wave front of the rotating source and for helpful comments that improved the manuscript.   

The work has been performed within the frames of the Science Projects of SPbSU 6.0.22.2010 and 6.38.669.2013.

 \end{document}